\documentclass[prl,twocolumn,showpacs,preprintnumbers,superscriptaddress,
amsmath,amssymb,tightenlines,epsfig]{revtex4}

\usepackage{bm}
\usepackage{graphicx}

\newcommand{\rv}{{\bf r}}

\newcommand{\beq}{\begin{equation}}
\newcommand{\eeq}{\end{equation}}
\newcommand{\bea}{\begin{eqnarray}}
\newcommand{\eea}{\end{eqnarray}}

\newcommand{\commentout}[1]{{}}

\begin{document}

\title{Energetically stable particle-like Skyrmions in a trapped Bose-Einstein condensate}
\author{C.M. Savage}
\affiliation{Department of Physics and Theoretical Physics, Australian National University, ACT 0200, Australia}
\email{craig.savage@anu.edu.au}
\author{J. Ruostekoski}
\affiliation{Department of Physical Sciences, University of
Hertfordshire, Hatfield, Herts, AL10 9AB, UK}
\email{j.ruostekoski@herts.ac.uk}

\begin{abstract}
We numerically show that a topologically nontrivial 3D Skyrmion can be energetically stable in a trapped two-component atomic Bose-Einstein condensate, for the parameters of $^{87}$Rb condensate experiments. The separate conservation of the two atomic species can stabilize the Skyrmion against shrinking to zero size, while drift of the Skyrmion due to the trap-induced density gradient can be prevented by rotation or by a laser potential. 
\end{abstract}

\date{\today}
\maketitle

Localized topological excitations that do not perturb the order parameter field at large distances from the particle, and which are characterized by a topologically invariant winding number, are well-known in nuclear and elementary particle physics \cite{SKY61,WIT83,BAT01,FAD97}. While their study in nuclear physics remains an experimental challenge, the recent experimental progress in atomic Bose-Einstein condensates (BECs) with internal spin degrees of freedom \cite{MAT99,LEA02} has raised the possibility of the existence of well-localized topological Skyrmions in atomic gases \cite{RUO01,Stoof,BAT02}. In this Letter we identify, and show how to overcome, the specific instabilities of Skyrmions in trapped two-species atomic BECs, and hence demonstrate their energetic stability under realistic experimental conditions.

Battye {\it et al.}\ \cite{BAT02} recently considered an infinite homogeneous two-species BEC, with constant total atom density. They showed that an energetically stable Skyrmion may exist as a result of phase separation of the two species, which suppresses the decay. These calculations were extended to non-constant total atom density, and to the trapping of one component \cite{BAT02b}. In this paper we show that in a harmonically trapped system there are additional instabilities, not considered in Refs.~\cite{BAT02,BAT02b}, which will play a crucial role in the experimental realization of Skyrmions in atomic BECs. Additional physical mechanisms, such as rotation or optical potentials, will be required for stability. We also show how density fluctuations, associated, e.g., with phonon emission \cite{adams}, are important in the Skyrmion decay process.

\begin{figure}[!b]\vspace{-0.5cm}
\includegraphics[width=\columnwidth]{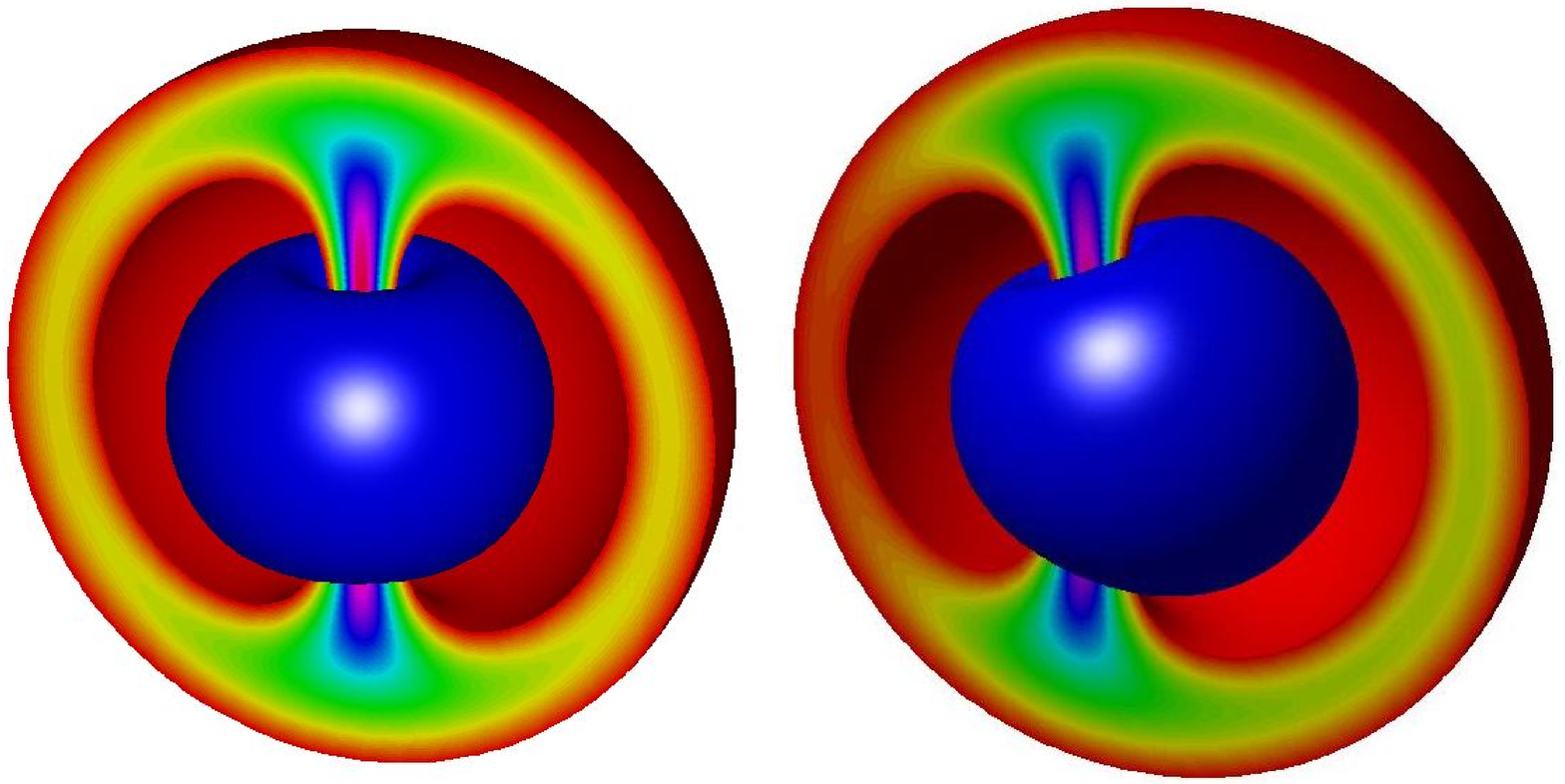}
\includegraphics[width= \columnwidth]{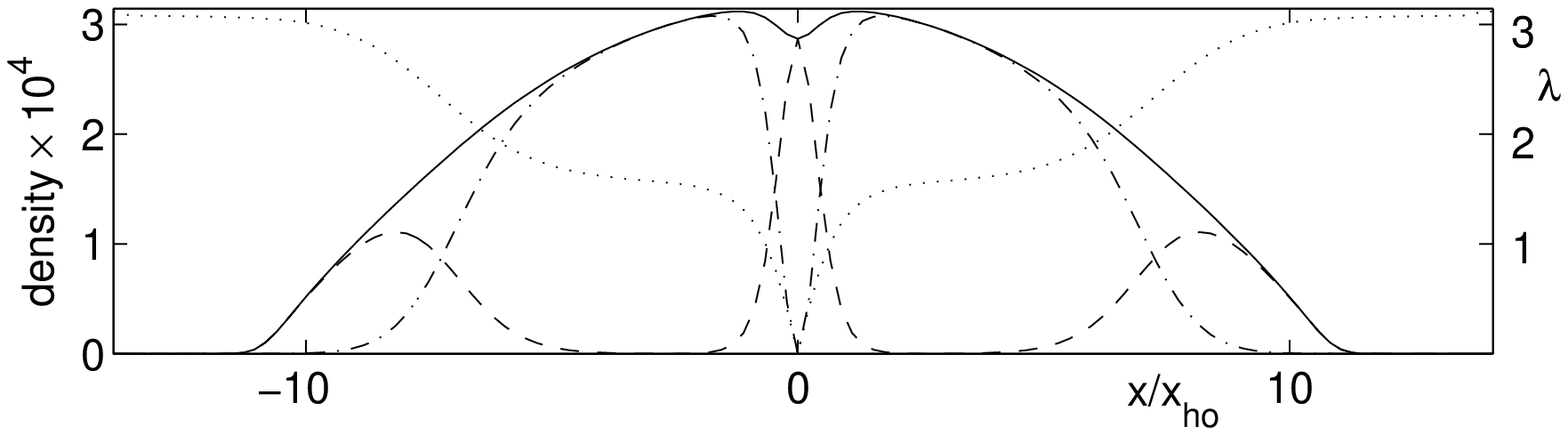}
\caption{Density and order parameter profiles for energetically stable trapped Skyrmions ($N_+ = N_- = 4.5 \times 10^6$). Top: 3D densities. The central (blue) torii are isosurfaces of $| \psi_- |^2$. Isosurfaces of $| \psi_+ |^2$ (red), are shown for $x < 0$: on the $y$-$z$ plane between the isosurface sections its density is indicated by a colormap from red (lowest) to purple (highest). Left: Stabilized by rotating  $\psi_-$ only, with angular velocity $0.1\omega$. Right: Stabilized by rotating the entire system with angular velocity $0.085\omega$. Bottom: 1D densities (left axis, units of $x_{ho}^{-3}$) and the order parameter profile $\lambda(x,0,0)$ (right axis, dotted line) for rotating  $\psi_-$ only. Solid line: $| \psi_+ |^2 + | \psi_- |^2$. Dashed line: $| \psi_+ |^2$. Dash-dot line: $| \psi_- |^2$. $\lambda(x,0,0)$ is extracted from the wave functions. The notch in $\lambda$ is due to the mean-field repulsion inflating the vortex ring core. $\lambda(\rv)$ is highly anisotropic, with $\lambda(0,0,z)$ qualitatively close to $2\arctan(|z|)$.}
\label{isosurface plot}
\end{figure}

There has been an explosion of interest in vortex and soliton experiments in atomic BECs \cite{ANG}, and we anticipate similar developments for other topological objects. Hence we identify the Skyrmion energetic stability criteria for the parameters of the JILA two-species $^{87}$Rb experiments \cite{MAT99}. We find a threshold frequency below 0.1$\omega$, when only one species is rotated, and a narrow window of rotation frequencies for the entire system around 0.085$\omega$. We numerically evaluate the stable configurations (Fig.~\ref{isosurface plot}) by minimizing the energy of the full 3D mean-field theory of coupled Gross-Pitaevskii equations (GPEs) and calculate the associated topological charges.

A Skyrmion is a topological particle-like soliton solution with a coreless 3D texture. Besides their intrinsic fundamental interest, Skyrmions have important applications in nuclear physics \cite{SKY61,WIT83,BAT01}, and analogous structures are postulated for early Universe cosmology \cite{DAV89}. Skyrmions are localized objects such that the order parameter field becomes uniform sufficiently far from the particle. As a result, we may enclose the Skyrmion by a sphere on which the field configuration has a constant asymptotic value. This sphere can be mapped to a single point in the order parameter space. Topologically, it indicates that the 3D physical region inside the sphere can be represented by a compact 3-sphere $S^3$ (a sphere in 4D) \cite{com2}. Here we consider a SU(2) order parameter space which topologically also corresponds to $S^3$. Then the mappings from the compactified 3D physical region to the order parameter space are represented by the field $U(\rv)$: $S^3\rightarrow SU(2)$. The crucial point is that having associated the enclosing sphere with a single point in $S^3$, we place the elements of the physical space into a correspondence with the elements of the compact order parameter space SU(2). The mappings from $S^3$ to SU(2) fall into topological classes, each characterized by an integer-valued, topologically invariant winding number $W$ \cite{SKY61,RUO01,BAT02}:
\beq
W={\epsilon_{\alpha\beta\nu}\over 24\pi^2}\int d^3 r\, {\rm Tr}[U(\partial_\alpha U^\dagger) U(\partial_\beta U^\dagger) U(\partial_\nu U^\dagger)]\,.
\label{win1}
\eeq
Here $\epsilon_{\alpha\beta\nu}$ denotes the completely antisymmetric tensor, the repeated indices are summed over, and $\partial_i$ represents the derivative with respect to the spatial coordinate $x_i$.  The topological charge describes how many times the set of points in physical space is `wrapped' over the order parameter space of SU(2) for a given element $U(\rv)$.

With cold atoms an SU(2) order parameter space is afforded most simply by a two-component BEC, whose interactions effectively fix the local value of the total density $|\psi_{+}(\rv)|^{2}+|\psi_{-}(\rv)|^{2} = \rho(\rv) $ of the two complex macroscopic wave functions $\psi _{+}$ and $\psi _{-}$ \cite{RUO01}. We may then express the two-component BEC as
\beq
\begin{pmatrix}
\psi_+(\rv) \\  \psi_-(\rv)
\end{pmatrix}
= \sqrt{\rho(\rv)} \, U^\dagger (\rv)
\begin{pmatrix}
1 \\ 0
\end{pmatrix} \,.
\eeq
We search for a topologically nontrivial solution for $U(\rv)$ with a nonvanishing winding number, determined by Eq.~(\ref{win1}). By minimizing the energy of the corresponding trapped BECs, we may investigate the {\it energetic stability} of the Skyrmion and find its equilibrium configuration. As an initial state for the numerical simulations we use $U(\rv)=\exp[i\lambda(\rv)\vec{\sigma}\cdot \hat{\rv}]$ \cite{com2b}, or a similar state displaced from the trap center, where $\sigma_i$ denote the Pauli spin matrices and $\hat{\rv}$ represents the unit radial vector. A direct substitution into Eq.~(\ref{win1}) yields $W=1$ for a monotonic function $\lambda$ with $\lambda({\bf 0})=0$ and $\lambda=\pi$ at the gas boundary. The corresponding BEC wave functions read:
\beq\label{skyrmsol}
\begin{pmatrix}
\psi_+(\rv) \\ \psi_-(\rv)
\end{pmatrix}
= \sqrt{\rho}
\begin{pmatrix}
\cos[\lambda(\rv)]-i\sin[\lambda(\rv)]\cos\theta \\
-i\sin[\lambda(\rv)]\sin\theta\exp(i\phi)
\end{pmatrix} \,.
\eeq
Here $(\lambda,\theta,\phi)$ can also be understood as the spherical angles of the 3-sphere. Then the boundary of the atom cloud corresponds to the pole $\lambda=\pi$ of the 3-sphere.

Due to the {\it topological stability} of the Skyrmion, any continuous deformation of Eq.~(\ref{skyrmsol}), without altering the asymptotic boundary values, is still a Skyrmion with $W=1$. Here, $\psi_-$ (the ``line component'') forms a vortex line with one unit of circulation around the core oriented along the $z$ axis (Fig.~\ref{isosurface plot}). Moreover, $\psi_+$ (the ``ring component'') forms a vortex ring with one unit of circulation around its core on the circle $z=0, r=\lambda^{-1}({\frac{\pi}{2}})$ \cite{RUO01}. The line component is confined in a toroidal region inside the ring component in the neighborhood of the ring core.

Despite its topological stability, the Skyrmion solution in several physical systems is energetically unstable against shrinking to zero size without additional stabilizing features. This can be understood by means of simple scaling arguments: if the kinetic energy density, or the order parameter `bending energy', is quadratic in the gradient of the order parameter, this energy density scales as $1/R^2$ with respect to the size $R$ of the Skyrmion. Because the Skyrmion occupies a volume proportional to $R^3$, the energy $E\propto R$ monotonically decreases with the size of the Skyrmion. In the Skyrme model, the stability is provided by an additional interaction term, with the energy density scaling as $1/R^4$~\cite{SKY61}.

It was proposed that, in a two-component BEC, the separate conservation of the species can effectively stabilize the Skyrmion in a homogeneous space against collapse to zero size \cite{BAT02}. In the regime of phase separation, with the scattering lengths satisfying $a_{++} a_{--} \lesssim a_{+-}^2$, the two species can strongly repel each other and the toroidal filling due to the line component can prevent the vortex ring from shrinking to zero radius \cite{RUO01,BAT02}. Consequently, a filled vortex ring, as opposed to an empty vortex ring, can be more stable against collapse. Moreover, in the Skyrmion the filling has one unit of angular momentum about the $z$ axis resulting in a $1/r^2$ centrifugal barrier that further prevents the shrinking. This should be contrasted to the Skyrmions in a ferromagnetic spin-1 BEC \cite{Stoof}, which are closely related to the Skyrmion excitations proposed in superfluid liquid $^3$He \cite{VOL77}, and can freely mix the atoms between the different spin components. As a result, no stabilizing features due to the atom number conservation exist in that case.

We next turn to numerical studies of the energetic stability of the Skyrmion using the full 3D mean-field theory of the coupled GPEs without additional simplifications:
\begin{equation}
i \hbar \dot\psi_i = \big( -{\frac{\hbar^2}{2m}} \nabla^2+V(\rv)+\sum_k \kappa_{ik} |\psi_k|^2
\big) \psi_i\,.  \label{gpe}
\end{equation}
Our simulations are performed for the parameters of the JILA two-species vortex experiment using perfectly overlapping isotropic trapping potentials for both states $V(\rv)=m\omega^2 r^2/2$, with $\omega = 2\pi \times 7.8$ Hz and harmonic oscillator length $x_{ho} \equiv (\hbar/m \omega)^{1/2}\simeq3.86 \mu$m \cite{MAT99}. In the interaction coefficients, $\kappa_{ij} \equiv 4\pi\hbar^2 a_{ij} N_i /m$, the number of atoms in each species is represented by $N_i$, $a_{ii}$ denotes the intraspecies, and $a_{ij}$ ($i\neq j$) the interspecies, scattering length. For the $|F=2, m_f =2 \rangle$ and $|1,1 \rangle$ spin states of $^{87}$Rb we have \cite{HALL98}: $a_{++}=5.67$nm, $a_{--}=5.34$nm, $a_{+-}=5.50$nm. The GPEs depend on the dimensionless interaction coefficients ${\kappa^\prime}_{ij} \equiv 4\pi N_i  a_{ij} / x_{ho}$. Hence, the dynamics is unchanged in any scaling of length and time, which does not change $N_i^2\omega$.

We found the ground states by imaginary time evolution of the GPEs (\ref{gpe}), using Runge-Kutta \cite{RK4IP} and split-step algorithms. 
At every time step, we separately normalized both wave functions to fix the atom number in each component. The numerical simulations were fully 3D, allowing for cylindrically asymmetric dynamics. The most demanding numerics was performed on a parallel multiprocessor supercomputer, using up to 32 processors. Spatial grids of $128^3$, $256^3$, and $512^3$ were used. For a typical spatial range of about $30 x_{ho}$ the corresponding grid spacings are respectively $0.23 x_{ho}$, $0.12 x_{ho}$, and $0.06 x_{ho}$. A typical imaginary time step was $0.0025/ \omega$.

As a cylindrically symmetric initial state we used the Skyrmion (\ref{skyrmsol}) with the Thomas-Fermi density profile \cite{com2b}, resulting in the winding number $W=1$. The winding number was calculated during the imaginary time evolution to determine the stability of the Skyrmion. For small BECs, with $N_T=10^4$ atoms ($N_T \equiv N_{+} + N_{-}$), the nonlinear repulsion between the two species was too weak to prevent the Skyrmion from shrinking for any $N_-/N_T$. Also the line component $\psi_-$ rapidly diffused to the boundary of the finite-size atomic cloud, altering the boundary conditions and resulting in a decreasing winding number $W<1$. Because the topological stability of the Skyrmion necessitates a well-defined asymptotic boundary condition with no phase variation, the Skyrmion was clearly lost. This decay mechanism is characteristic of trapped BECs and cannot occur in homogeneous systems, which implicitly assume large $N_+/N_T$.

\begin{figure}
\includegraphics[width=\columnwidth]{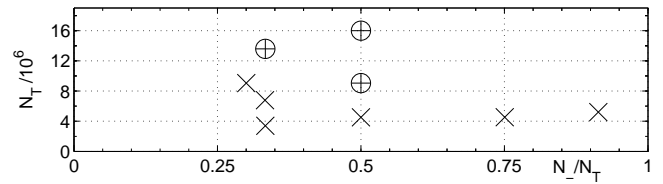}
\caption{Skymion stability diagram. Total atom number $N_T$ vs. the fraction in the line component  $N_{-}/N_T$. $\times$ indicates instability. $\oplus$ indicates stability of the cylindrically symmetric state against shrinkage, which may be stabilised against drift as discussed in the text.}
\label{stability diagram}
\end{figure}

\begin{figure}
\includegraphics[width=\columnwidth]{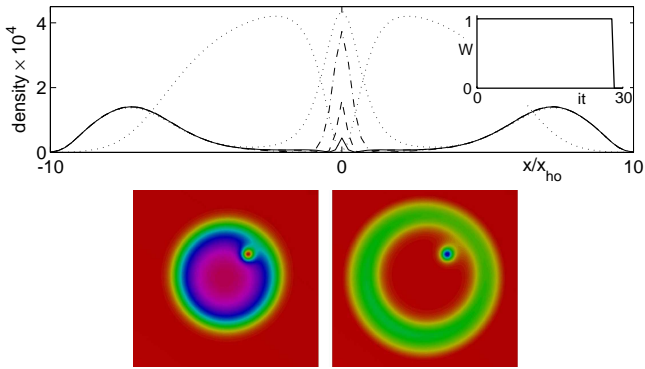}
\caption{Decaying Skyrmion's imaginary time evolution. Top: Densities (units of $x_{ho}^{-3}$) along the $x$ axis.  $| \psi_+ |^2$: $it=20$ (dash-dot), $27.75$ (dashed), $28$ (solid), $50$ (dotted). The dotted curve which is zero at $x=0$ is $| \psi_- |^2$ at $it=50$: for other times it is qualitatively similar. $W=0$ by $it=28.25$, and $it=50$ is the stationary state. $N_+ = N_- = 2.3 \times 10^6$. Inset: Winding number W versus imaginary time. Bottom: Densities on the $z=0$ plane, left $| \psi_- |^2$, right $| \psi_+ |^2$. Since there is no stabilisation, the line vortex core (red circle) has moved towards the boundary and the ring vortex (surrounding the blue circle) is about to collapse. Colormap and parameters as for Fig.~1.}
\label{decay}
\end{figure}

For large BECs with $N_T>10^6$, represented in Fig.~\ref{stability diagram}, the same instability mechanism was observed with  a very large fraction of line component $N_-/N_T \gtrsim 0.75$. In other cases displayed in Fig.~\ref{stability diagram}, the winding number typically evaluated to either 0 or 1 to better than 1\%, with $\lambda=\pi$ well preserved at the boundary. A drop from $W=1$ to $W=0$ indicated instability of the Skyrmion against shrinking, see inset to Fig.~\ref{decay}. The Skyrmion collapse via shrinking occurred when the high density central part of the ring component, which passes through the line component vortex core, pinched off. The total density varies significantly at the trap center when the vortex ring core collapses, emphasizing the importance of density fluctuations in the decay process, see Fig.~\ref{decay} (top). With sufficiently large BECs, even for the collapsed Skyrmions with no vortex ring, the local energetic minimum can correspond to line component trapped inside a toroidal region with $\psi_+$ forming an atom-optical confinement.

For sufficiently large total and line component atom numbers, the nonlinear repulsion between the two species was strong enough to inhibit the collapse of the vortex ring, stabilizing the Skyrmion against shrinking for cylindrically symmetric initial states (Fig.~\ref{stability diagram}). However, due to the inhomogeneous potential, the Skyrmion was still unstable with respect to drift towards the edge of the BEC where its energy is lower. This was found using cylindrically asymmetric initial states, emphasizing the importance of the full 3D simulation. The drift occurred with the vortex line moving towards the boundary. Once the vortex line drifted to a low-density region, the nonlinear repulsion was no longer strong enough to prevent the shrinking of the Skyrmion and the collapse occurred as described above, see Fig.~\ref{decay} (bottom). The drift reduces the total angular momentum of the atoms indicating an energetic instability. Physically, this results from dissipation, e.g., due to thermal atoms.

We have investigated several mechanisms by which Skyrmions might be stabilized against the vortex drift instability. Perhaps the simplest is to rotate the line vortex component about the $z$ axis, above its critical angular velocity, thus stabilizing the vortex. This should always work for cylindrically symmetric traps in which the symmetric density prevents the rotation coupling into the other BEC component.  For the case of Fig.~\ref{isosurface plot} a line component angular velocity of $0.1 \omega$ stabilised the Skyrmion.

Rotation of both species introduced a new instability mechanism. For sufficiently rapid rotations the line component again reached the boundary, resulting in $W<1$. At the same time the ring vortex accomodated the rotation by twisting and finally breaking at the gas boundary. Nevertheless, for the case of Fig.~\ref{isosurface plot}, we found a small range of angular velocities around $0.085 \omega$ for which the Skyrmion was stable against this, and against line vortex drift. It stabilized off the rotation axis, indicating an equilibrium between the lower energy of the line vortex and the higher energy of the ring component threading through it with increased distance from the trap center. The breaking of cylindrical symmetry implies a family of degenerate Skyrmions parametrized by the polar angle~$\phi$.

Another method for stabilizing the Skyrmion is to inhibit the drift towards low-density regions by creating a positive density gradient around the vortex line. This can be implemented with a blue-detuned Gaussian laser beam along the $z$ axis \cite{blue beam}, providing a cylindrically symmetric repulsive Gaussian dipole potential perpendicular to the $z$ axis. The total potential then has the ``Mexican hat'' form:
$V = m\omega^2 r^2 /2 + V_0  \exp[-2 (x^2+y^2)/w^2 ]$. Setting $w = 7 x_{ho}$ and $V_0 = 25 \hbar \omega$, corresponding to about twice the width of the vortex line core, successfully stabilized a Skyrmion with $N_+ = N_- = 8 \times 10^6$.

Proving numerically that a Skyrmion is stable is difficult. The essential requirement is that it be stationary under imaginary time evolution. This was determined by plotting the density on 1D sections through the system, and the phase variation on 2D slice planes. In unstable cases these would evolve until the Skyrmion decayed. When they became asymptotically constant as a function of imaginary time the Skyrmion was judged to be stable. Convergence was particularly slow, and hence difficult to establish, when both components were rotated.

The Skyrmion can be created using electromagnetic fields to imprint topological phase singularities on the matter field while changing the internal state of the atoms, as proposed in Ref.~\cite{RUO01}. In particular, a vortex ring can be engineered in a controlled way with an appropriate phase-coherent superposition of three orthogonal standing waves. Due to dissipation, the prepared Skyrmions relax to the ground states calculated here.

Different spin textures are often referred to as ``Skyrmions'', even when they are not characterized by the topological invariant (\ref{win1}). It is important to emphasize that the Skyrmions studied in this paper are fundamentally different from the nonsingular Anderson-Toulouse or Mermin-Ho vortices \cite{AND77}, frequently also called Skyrmions \cite{Ho}.
Although these vortices may be energetically stable in a rotating trap \cite{MIZ02}, they are topologically trivial in the 3D ferromagnetic spin-1 BEC and, analogously to the vortex lines with two units of circulation, they can be continuously deformed to a topologically trivial uniform spin distribution.

\acknowledgments
{This research was supported by the EPSRC, the Australian Research Council, and the Australian Partnership for Advanced Computing.}


\begin{thebibliography}{99}
\bibitem{com2} All such configurations can be associated with a class of mappings that comprises an element of the third homotopy group $\pi_3$. Two such configurations can be deformed into one another in a singularity-free space without altering the asymptotic values, if and only if they are represented by the same element of $\pi_3$. Here $\pi_3[SU(2)]=Z$ indicates that topologically different configurations can be characterized by an integer-valued winding number.
\bibitem{com2b} Without the loss of generality, we may use as an initial state $\lambda=\lambda(r)$, since the spherical symmetry of $\lambda$ is not conserved in the imaginary time evolution.
\end{thebibliography}
\end{document}